\documentclass[conference]{IEEEtran}
\IEEEoverridecommandlockouts

\usepackage[utf8]{inputenc}

\usepackage{amsmath,amssymb,amsfonts}
\usepackage{cite}
\usepackage{enumitem}
\usepackage[hang,flushmargin]{footmisc}
\usepackage{graphicx}
\usepackage{mathptmx}
\usepackage{textcomp}
\usepackage{xcolor}

\newcommand{\red}[1]{{\color{red}#1}}

\newcommand{\m}[1]{{\color{black}#1}} % ELISA
\renewcommand{\IEEEauthorrefmark}[1]{{\textsuperscript{#1}}}

\begin{document}

\title{Expanding Frontiers: Settling an Understanding 
%on 
\m{of} Systems-of-Information Systems}

\author{
\IEEEauthorblockN{Valdemar Vicente 
Graciano Neto\IEEEauthorrefmark{1}, Bruno 
%Gabriel Araújo 
Lebtag\IEEEauthorrefmark{1}, Paulo Gabriel 
Teixeira\IEEEauthorrefmark{1}, Priscilla 
%Elizabeth Pereira 
Batista\IEEEauthorrefmark{1}, Vinícius 
%Carvalho 
Lopes\IEEEauthorrefmark{1}, \\Jamal El-Hachem\IEEEauthorrefmark{2}, Jérémy Buisson\IEEEauthorrefmark{2}, Flavio Oquendo\IEEEauthorrefmark{2}, Juliana Fernandes\IEEEauthorrefmark{3}, Francisco Ferreira\IEEEauthorrefmark{4}, Rodrigo 
%Pereira dos 
Santos\IEEEauthorrefmark{4}, \\Davi Viana\IEEEauthorrefmark{5}, Everton Cavalcante\IEEEauthorrefmark{6}, Mohamad Kassab\IEEEauthorrefmark{7}, Ahmad Mohsin\IEEEauthorrefmark{8}, Roberto Oliveira\IEEEauthorrefmark{9}, Vânia Neves\IEEEauthorrefmark{10}, \\
\m{Maria Istela Cagnin\IEEEauthorrefmark{11}} and Elisa Nakagawa\IEEEauthorrefmark{12}}\smallskip

\IEEEauthorblockA{
\IEEEauthorrefmark{1}Federal University of Goiás, Goiânia, Brazil}
\IEEEauthorblockA{\IEEEauthorrefmark{2}IRISA, University of South Brittany, Vannes, France}
\IEEEauthorblockA{\IEEEauthorrefmark{3}Federal Institute of Piauí, Campo Maior, Brazil}
\IEEEauthorblockA{\IEEEauthorrefmark{4}Federal University of the State of Rio de Janeiro, Rio de Janeiro, Brazil}
\IEEEauthorblockA{\IEEEauthorrefmark{5}Federal University of Maranhão, São Luis, Brazil}
\IEEEauthorblockA{\IEEEauthorrefmark{6}Federal University of Rio Grande do Norte, Natal, Brazil}
\IEEEauthorblockA{\IEEEauthorrefmark{7}The Pennsylvania State University, Malvern, USA}
\IEEEauthorblockA{\IEEEauthorrefmark{8}Edith Cowan University, Perth, Australia}
\IEEEauthorblockA{\IEEEauthorrefmark{9}State University of Goiás, Posse, Brazil}
\IEEEauthorblockA{\IEEEauthorrefmark{10}Fluminense Federal University, Niterói, Brazil}
\IEEEauthorblockA{\IEEEauthorrefmark{11}\m{Federal University of Mato Grosso do Sul, Campo Grande, Brazil}}
\IEEEauthorblockA{\IEEEauthorrefmark{12}University of São Paulo, São Carlos, Brazil\smallskip\\
Email: valdemarneto@ufg.br\vspace*{-0.5cm}}
}

\maketitle

% If you want to contribute, these are the duties still required to be performed:
% 1. Draw Figure 1 according to the text available in Section III (a sequence diagram);
% 2. Review and improve English writing;
% 3. Identify confusing or non-clear text excerpts;
% 4. Identify eventual conceptual inconsistencies;
% 5. Adding missing information, citations (references) and text complements (on mKAOS, for instance);
% 6. Adding your own name and institution in the paper heading (order will be defined later);
% 7. Other improvements that I did not glimpse. 

\begin{abstract}
%Systems-of-Systems (SoS) have been consolidated as a research area.
\m{System-of-Systems (SoS) 
has consolidated itself as a special type of software-intensive systems.} As such, %SoS subdomains 
\m{subtypes of SoS} have also emerged, such as Cyber-Physical SoS (CPSoS\m{) that are}
%, 
formed essentially 
%by 
\m{of} cyber-physical constituent systems
%) 
and Systems-of-Information Systems (SoIS\m{) that}
%, which 
contain 
%software-intensive Information Systems among the
\m{information systems as their}
constituents\m{.}
%). 
In contrast to CPSoS
%, which 
%has 
\m{that have} been investigated and covered in the specialized literature, SoIS still 
%lacks some
\m{lack} critical discussion about 
%its 
\m{their} fundamentals. The main contribution of this paper is %investigating 
\m{to present} those fundamentals to set an understanding 
%on 
\m{of} SoIS. By offering a discussion and examining 
%some 
literature cases, we draw an essential settlement on SoIS definition, basics, and practical implications. The discussion herein presented results from research conducted on SoIS over the past years in interinstitutional and multinational research collaborations. The knowledge gathered in this paper arises from several scientific discussion meetings among the authors. As a result, we aim to contribute to the %state-of-the-art
\m{state of the art of} 
%by converging the 
SoIS 
%fundamentals 
besides paving \m{the} research avenues for the forthcoming years.
\end{abstract}


\section{Introduction}
\label{s:introduction}
Systems-of-Systems (SoS\footnote{We will herein interchangeably use SoS acronym to express both singular and plural forms (System-of-Systems and Systems-of-Systems).}) have become popular in scientific vehicles. Workshops\m{,} %and thematic sessions have been proposed
\m{technical sessions in large events, and other foruns have occured addressing SoS} %in several fields 
%\footnote{Examples include the SESoS/WDES 2021, i.e., the Joint 9th edition of the International Workshop on Software Engineering for Systems-of-Systems (SESoS) and the 15th edition of the Workshop on Distributed Software Development, Software Ecosystems and Systems-of-Systems (WDES), co-located with the 2021 43rd International Conference on Software Engineering (ICSE), SiSoS at ACM SAC track, and the Track on Model Engineering for System of Systems held together with The 31st European Modelling \& Simulation Symposium (EMSS) in 2019 and 2020.}
and even a tertiary study 
%has been 
\m{was} already conducted 
%on the subject 
\cite{Cadavid2020}
%, which reveals a 
\m{revealing the} growing maturity 
%in 
\m{of} the area. %Moreover, the investments in smart cities, which are remarkable instances of SoS, have reached nearly \$124 billion mark worldwide\footnote{https://www.smartcitiesdive.com/news/idc-worldwide-smart-city-spending-124B-2020/572352/}. These facts indicate how prominent and promising the SoS research area have become.

Apart from those advances, technological and theoretical gaps still remain. We perceive important theoretical gaps, particularly in subfields that have emerged, such as Systems-of-Information Systems (SoIS\footnote{We will herein interchangeably use SoIS acronym to express both singular and plural forms (System-of-Information Systems and Systems-of-Information Systems).}) and Cyber-Physical Systems-of-Systems (CPSoS). CPSoS, i.e., the SoS formed by Cyber-Physical Systems (CPS), are the main topic of SoSE conference in 2021 and currently involve more than 400 studies retrieved in Google Scholar\footnote{Results of a search for \texttt{"cyber-physical systems-of-systems"}: https://bit.ly/3a7NtVn} at the moment of creating this paper, revealing that the area is being consolidated over the years. In turn, a 
%recent 
systematic mapping study %\cite{Teixeira2019} 
revealed that
%, up to 2018, 
only 25 studies referring to SoIS had been published \m{until 2018 \cite{Teixeira2019}}. The number of studies and the findings brought by the mapping could be seen as evidence of the need for further investigation in \m{the} SoIS area. Particularly, topics such as specific notations for SoIS modeling that could support its specificities and the respective engineering methods and tools are among the advances needed, once an imprecise or ambiguous  specification could potentially lead to defects, malfuntion and potential losses.

Due to the relative novelty and scarce literature on SoIS, researchers still ask questions such as: 
\begin{enumerate}[label={Q\arabic*}:,leftmargin=*,after={\medskip}]
    \item \textit{What is SoIS?}
    \item \textit{What are the differences between SoS and SoIS?}
    \item \textit{What are the specific characteristics for an SoS to be considered as a SoIS?}
    \item \textit{Why is SoIS necessary as a particular type of system?}
    \item \textit{What are the engineering needs to be raised by SoIS as a specific SoS subtype?}
\end{enumerate}

We are aware that SoIS (as CPSoS) are not new types of systems, but SoS subtypes. Every SoIS or CPSoS is actually an SoS with \textit{additional characteristics}. %The predominance of specific types of software-intensive systems, such as Information Systems (IS) and CPS, changes the way that SoS engineering is done, potentially requiring new notations, methods and techniques, and consequently motivating the enunciation of new subclasses to study their specific characteristics and propose tools that are more specific and suitable for their engineering. 
While CPSoS have been broadly explored over the past years, with well-established discussion about their definitions and fundamentals
\cite{engell2015core}, SoIS still lacks this type of discussion. We noticed that SoIS impose particular engineering challenges that the state of the art on SoS had not necessarily coped with. Hence, advances on SoIS fundamentals and upperlying methods and tools are needed\m{; otherwise,}
%, otherwise 
future SoIS specifications can be imprecise, leading to low\m{-}quality SoIS and potential losses and/or injuries due to malfunction. 

This paper focuses on an investigation on the SoIS fundamentals. The main contribution is 
%answering 
\m{to answer} the aforementioned questions by (i) presenting evidence found in the literature and (ii) building an understanding 
%on 
\m{of} SoIS over the knowledge acquired over the past years of research in interinstitutional and multinational collaborations. Content of this paper reflects a consensual view involved in the discussions over the years and the knowledge built in a research network made up of 
%eleven 
\red{12} different institutions 
%and 
\m{in} four different countries. We expect to better disseminate the SoIS research area and contribute to the Systems-of-Systems Engineering community's advances.

Given the elucidative nature of this paper, its organization may be different from others since we use the literature to both support the background and answer the raised questions. Section~\ref{s:relatedWork} presents the background with essential concepts, besides SoIS examples in literature and advances from the state of the art. Section~\ref{s:understanding} establishes an understanding for SoIS by answering the raised questions. Section
~\ref{s:discussion} brings further discussions derived from the established foundations. Section~\ref{s:gaps-challenges} lists some research challenges and gaps. Finally, Section~\ref{s:finalRemarks} concludes the paper with final remarks.

\section{Background}
\label{s:relatedWork}

%SoS can be formed by several types of software-intensive constituent systems\footnote{Herein, we do not address SoS formed by systems that do not involve software, such as pure mechanical or analogic systems.} \cite{maier_architecting_1998,Cadavid2020}, i.e., systems whose life cycle is directly impacted by software, from its conception to deployment and maintenance, and that has software as an inherent part \cite{iso42010}. Examples of software-intensive systems include software-intensive IS, embedded systems, robotic systems, automobile systems, industrial systems, and CPS\footnote{For our discussions, we adopt CPS as a comparison parameter since it is the most established SoS subtype.}. Except for IS, several systems considered software-intensive are usually named as such because, besides software, they hold a physical (i.e., electrical, mechanical, robotic, hydraulic) counterpart that allows them to perform their activities and act on the environment.  

%Ahmad: (Information Systems Defintion) An Information System consists of software, hardware and people with the capability to store, process and filter useful information for decision making. Examples include decision support systems, automation systems, management information systems and emergency management systems ?

SoIS are understood as SoS subtypes that necessarily involve software-intensive 
information systems (IS)\m{,}
%\footnote{Henceforth, we use IS acronym to denote %\textit{software-intensive} Information Systems.}, 
i.e., IS whose life cycle is directly impacted by software, from its conception to deployment and maintenance, and that has software as an inherent part \cite{iso42010}. A single IS is often materialized as a software-based set of associated components deployed in a hardware and operated by humans  that collect (or retrieve), process, store, and distribute information \cite{tomicic2012strategies,Laudon:2015}. As the IS definition is quite broad and could include several systems\m{,} such as an autonomous car or a smartwatch, our matter of investigation here are those IS that do not involve \m{only} 
%a 
physical counterpart\m{s}, such as drones or smart sensors (that we consider as CPS). Instances of IS 
%include 
\m{can be} decision support systems, enterprise systems, public finance systems, and social networks. This is important to remark since CPSoS and other SoS potential subtypes do not necessarily involve IS as constituents. For instance, a Flood Monitoring SoS \cite{Oquendo16a, Oquendo16b} and Smart Parking SoS \cite{Delecolle20} are often CPSoS since they majorly (or even totally) involve CPS, i.e., independent systems characterized by an extensive number of physical devices (e.g., sensors, controllers, etc.) and cyber components (software counterparts employed to collect data from sensors, act on the environment, monitor, and manage the underlying infrastructure) \cite{ELSHENAWY2018575}. In turn, smart cities or their inner sub-SoS can be considered SoIS \cite{neto2017smart,neto2017towards,Fernandes2019,Fernandes2020}, once they potentially involve IS to manage the city infrastructure, which can dramatically change the way the emergent behaviors are planned and designed. 

\textbf{
%Major (Major dá impressão que são os mais importantes do mundo, o que não é bem verdade).
\m{Some}
SoIS initiatives.} Some examples of SoIS exist in the literature.  
%, such as Neves et al. \cite{Neves2020} and Majd and Marie-Hélène \cite{majd2017system}, that deal with Educational SoIS. 
Neves et al. \cite{Neves2020} investigate the need for interoperating several different educational IS (essentially virtual learning environments, known as VLE) to (i) complement each other regarding the provided functionalities, (ii) abstract away the need of using several different VLE from users, and (iii) allow users to share all the functionalities offered by the constituents that form the SoIS. This SoIS is essentially formed by several IS, not involving other types of constituents. Majd and Marie-Hélène \cite{majd2017system} also worked on educational SoIS. They modeled and developed the MEMORAeSoIS as a support for the learning ecosystem. The aim was to facilitate resource management in a SoIS, combining resources from several IS
%, 
and managing them within the leader system. This resulted in an added-value that would not be present if they were operating separately.

\begin{figure*}[!ht]
    \centering
	\includegraphics[width=2\columnwidth]{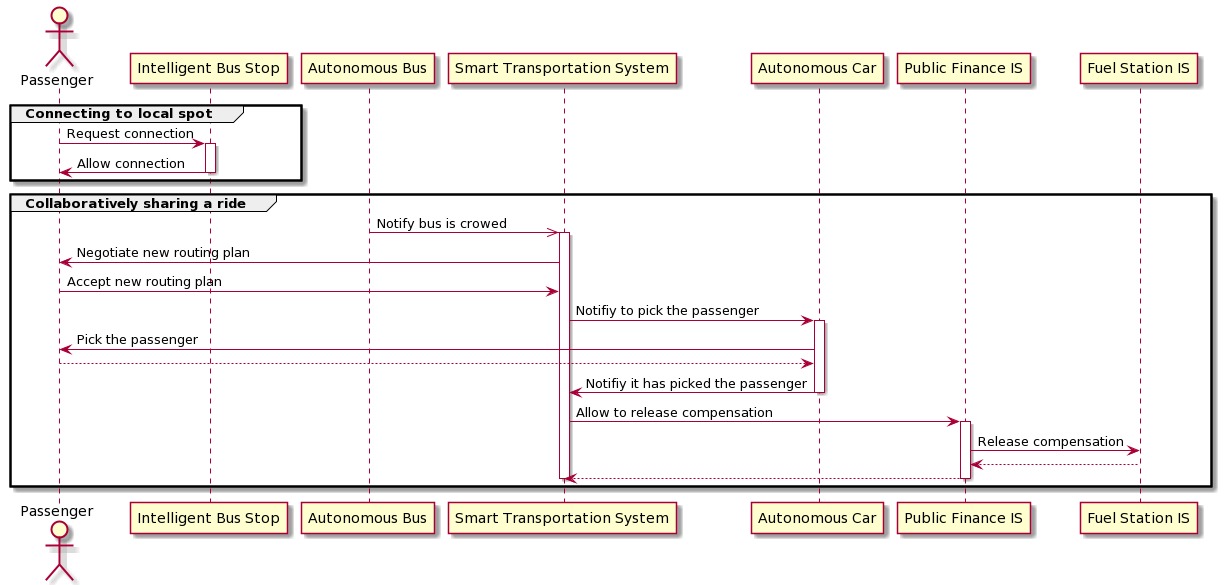}
	\caption{UML sequence diagram representing the message exchange among constituents in a smart transportation system of a smart city.}
	\label{fig:sequence-diagram}
\end{figure*}

In another context, Graciano Neto et al. \cite{neto2017towards} deal with smart cities as an example of SoIS. The author\m{s} 
%and colleagues 
motivate a scenario by analyzing a smart city. Several constituents are involved in a smart city SoIS, such as (i) autonomous cars and buses, (ii) intelligent bus stops, (iii) public finances IS, (iv) fuel station IS, and (v) the smart transportation system itself. %\red{(Foi dito no primeiro parágrafo desta seção que os SoIS que será abordado neste paper são aqueles que não tem partes físicas. Esse texto em magenta diz que esse exemplo de SoIS tem essas partes físicas. E daí parece que não estamos sendo muito coerentes.)}
Those systems then establish interoperability links to achieve a specific goal: \textit{enabling efficient transportation for population}. Fig. \ref{fig:sequence-diagram} illustrates a scenario in which buses are too crowded. As such, the buses communicate with the intelligent bus stops and autonomous cars so that passengers can be reallocated to arrive in their destination accordingly. Thus, passengers in that specific bus stop have their mobiles connected to a local spot. The bus stop, once received the crowding alert from the bus, notifies the passengers they will be reallocated to autonomous cars that are available for a ride because drivers and passengers share close destinations. Once a driver gives the ride to a passenger, s/he receives a discount in fuel stations readily authorized due to interoperability among the smart transportation system (formed by autonomous cars, buses, and bus stops), the public finance IS, and the fuel station IS.

% a UML sequence diagram or BPMN diagram
%Particularly in the second example, the emergent behavior can be expressed using an UML sequence diagram, as depicted in Figure \ref{fig:sequence-diagram}. The figure presents the two scenarios previously described: (i) the passangers connecting to the hub spot and (ii) the smart city SoIS sharing rides. However, an UML sequence diagram does not natively support the representation of the multiple systems and organizations involved in a SoIS. Both diagrams would represent a constituent archetype (i.e., an abstract illustration of some of the constituents mentioned) rather than all the constituents involved. UML and SysML are prepared to represent only one system (not multiple systems). When representing multiple systems, a single architectural change (e.g., a system that joins or leaves the SoIS) makes the documentation automatically outdated. Given those examples, in the next section, we discuss the foundations for SoIS collected from the literature and elaborated over the years.

\textbf{SoIS advances and state of the art.} Besides SoIS examples being described in the literature, other conceptual advances have also been proposed, such as a systematic mapping study on SoIS \cite{Teixeira2019}, a conceptual model for SoIS \cite{Fernandes2019}, a study on factors that influence interoperability in SoIS \cite{Fernandes2020} and a prototype of a SoIS essentially formed by several virtual learning environment systems \cite{Neves2020}.%, and mapping on the relation between SoS and business processes \cite{santos2020business}, which helps us to understand how business processes have been addressed in SoS so that we could also use those advances. 

%Another vital concern refers to the specification of goals (previously known in SoS domain as \textit{missions}), i.e.,  activities that are partitioned into smaller operational tasks assigned to specific constituent IS according to a matching of their capabilities to realize the planned emergent behaviors \cite{gracianoNeto2018SBSI}. The multiple institutions involved as a consequence of the managerial independence of constituents and the SoS/SoIS inherent evolutionary architecture imposes the need to . As such, advances in simulation and executable models can be welcome \cite{Lebtag2020}. Given that panorama, the next section distills some examples of SoIS from the literature, discussing their main characteristics.

% A SoIS is then formed by one or more software-intensive, operationally, and managerially independent IS interoperating with other systems to achieve common goals \cite{saleh2015information,Teixeira2019}. 

\section{Establishing an Understanding for SoIS}
\label{s:understanding}
Although all that knowledge \m{on SoIS} has already been disseminated in several vehicles, the questions raised in Section~\ref{s:introduction} were not yet readily answered. We then proceed with a deeper discussion to answer those questions. 

\subsection
%*{Q1: What is SoIS?}
{\textbf{Q1: What is SoIS?}}

Teixeira et al. \cite{Teixeira2019} identified two main research groups in SoIS\m{: the group of}
%, ours and 
Saleh, Abel\m{,} and colleagues \cite{saleh2015information} \m{and ours}. Our group claims the fact that SoIS requires distinct engineering techniques\m{, while the other}
%. The latter 
group faces SoIS as an SoS that behaves as an IS, 
%which does not preclude 
\m{not precluding} that 
%it 
\m{SoIS} require
%s 
distinctive engineering techniques. %Moreover, in our perspective, IS is a branch that gathers computer science and Business, different from abroad The French group considers the SoIS as a SoS that behaves as an IS, whilst our group understands that SoS could indeed fit the IS definition; however, we make this distinction clear (between IS and SoIS), once the engineering techniques for a SoIS can be dramatically different when compared to an IS and those differences (colaborating independent systems, managerial and operational independence, and dynamic architecture) allow the development of specific techniques and tools specific for SoS and SoIS.

Given the differences between the \m{conceptualization of SoIS,} 
%SoIS conceptions, 
our group 
%has proposed 
\m{proposes} the following definition \m{for SoIS as a result of} 
%as 
a compilation 
%from 
\m{of} the literature and
%as a result of 
a systematic mapping study \cite{Teixeira2019}: \textit{\m{``}A System-of-Information Systems (SoIS) is a specific 
%class 
\m{type} of SoS oriented to business processes in which \m{the} constituent systems include information systems that interoperate among 
%themselves 
\m{them} and belong to different %organizations/institutions/entities (OBS: tirei isso para passar a fazer sentido colocar uma outra definição em seguida.)
\m{organizations}.\m{''}} From 
%that 
\m{this} definition, we can discuss some important concerns. \textit{Oriented to business processes} means that, essentially, SoIS goals are drawn as well-established information flows among constituents, and the activities (sub-goals) performed by the constituents are interdependent, i.e., there is a specific sequence that should be followed to achieve the goal realization. \textit{Constituent systems include 
IS} means that other types of systems can also be involved in the SoIS, but at least one IS should be a constituent so that such SoS can be considered a SoIS. \textit{Belong to different organizations} means that 
%we aim to deal with SoIS  
\m{the SoIS are} formed by constituents with managerial independence.

In this context, herein the concept of 
%organization
\m{\textit{organization}}
can be relative for SoS/SoIS. For instance, in the Educational SoIS described by Neves et al. \cite{Neves2020}, the multiple VLE were deployed at the same university. This happened because the institution politics allowed (i) different departments to adopt their preferred VLE, (ii) professors adopt other VLE, and (iii) the university itself had a major VLE required to be used by everyone. However, each one was managed by a different department, which represents the constituents' managerial independence. Therefore, a SoIS can exist within the same organization, but different inner entities can manage constituents. To be more inclusive about scope \m{of a SoIS}, we deliver the following definition:\medskip

%\noindent\textbf{\textit{A SoIS is a specific type of SoS in which the constituents set include IS that interoperate with other constituents to achieve goals.}}\medskip

\noindent\textit{\m{``A SoIS is a specific type of SoS in which the \m{set of constituent systems} 
%constituents set 
include IS that interoperate with other constituents to achieve goals.''}}\medskip

We remark that SoIS goals are the realization of planned emergent behaviors and they are achieved as the result of the interoperation (information exchange) among the constituents. A necessary corollary is that, due to the inherent characteristics of SoS, the \textit{information exchange} among constituents will be frequently expressed as a \textit{business process} that is \textit{inter-organizational/\m{inter-entities} and flexible}.

%\subsection*{Q2: What are the differences between SoS and SoIS?\\ Q3\m{:} What are the specific characteristics for an SoS to be considered a SoIS?}

\subsection {\textbf{\m{Q2: What are the differences between SoS and SoIS? and Q3\m{:} What are the specific characteristics for an SoS to be considered a SoIS?}}}

%Every 
\m{As a} SoIS is an SoS\m{, it}
%. Therefore, every SoIS 
inherits the set of five characteristics that define SoS \cite{maier_architecting_1998}: \textit{operational independence}, \textit{managerial independence}, \textit{emergent behavior}, \textit{evolutionary development}, and \textit{distribution}. We argue that, for being 
%a SoIS, 
\m{an SoS,} 
%an SoS 
\m{a SoIS} should additionally hold the following characteristics:
%A consequence of operational independence of constituents and autonomy \cite{boardmansauser:2006} is the evolutionary (or dynamic) architecture \cite{Oquendo16a,Manzano2020}, i.e., the SoS (and the SoIS) architecture is continuously evolving due to the constituents that join or leave the SoS/SoIS at runtime. We remark that the SoS/SoIS architecture is as dynamic as the \textit{degree of independence} exhibited by the SoS/SoIS constituents \cite{Fernandes2019}. Hence, the changes cannot be so frequent, even being allowed.

\begin{itemize}
    
\item \textbf{Characteristic 1: IS Presence.} At least one IS should be
%involved 
\m{present} in the 
%SoS constituents set to be considered as a SoIS. 
\m{set of constituents.} From our experience, more than one IS is frequently involved in a SoIS. This means that the SoIS nature depends on the nature of the constituents that contribute to the SoIS.% as opposed to the nature of the services they provide.

\item \textbf{Characteristic 2: Goals are expressed as flexible and inter-organizational business processes.} SoIS are also concerned with the flow of information and knowledge among different IS. As stated by the definition, the SoIS goals are achieved by means of interoperation among constituents. Such an interoperation is materialized in information exchanges, task processing, and service invocation among them. This interoperability leads to the establishment of business processes that characterize the data exchange among constituents to achieve a goal\m{, as also found in a deep investigation that we performed on the relationship between SoS and Business Processes \cite{Cagnin2021b}}. We understand that business processes exist to support a set of organizational goals and there are, at least, two types of emergent behaviors: \m{(}i) the planned emergent behavior, which supports pre-established business process\m{;}
%, 
and \m{(}ii) unexpected emergent behavior, which can be beneficial or not to the SoIS. If the unexpected emergent behavior brings benefits, it can raise opportunities to create new business processes for SoIS. However, we highlight both result from interoperation among IS. As the architecture of a SoIS can be dynamic and constituents are managerially independent, goals should be expressed as business processes that should be splitted into smaller goals to be assigned to the involved constituents. However, since 
%a 
constituent\m{s} can leave\m{, enter, or be replaced in} a SoIS at runtime, the business process should then be\m{:} (i) flexible, i.e., the flows and activities should be allowed to change at runtime\m{;} and (ii) inter-organizational, i.e., encompass multiple organizations. A business process 
%preclude subgoals 
\m{establishes the} interdependence \m{among subgoals} and 
%it is an established 
sequence\m{s} of interdependent activities to achieve 
%a goal. 
\m{the goals.} If 
%the 
\m{a given} SoIS goal is expressed as a business process, then the subgoals should be performed in a given order\m{; otherwise,}
%, otherwise 
the effect would not be the intended one. This is important to remark because this is not necessarily true for other types of SoS. For instance, in a Flood Monitoring SoS, the emergent behavior (a flood alert) is usually achieved due to a pure message forwarding between the smart sensors until the gateway that processes the information and eventually triggers the flood alert. Hence, the goal is not established as a business process, but only as a message forwarding mechanism.

\end{itemize}

We claim that Characteristic 1 is mandatory for a SoS to be considered as a SoIS while Characteristic 2 is essentially a consequence of the presence of multiple organizations\m{/entities} involved in the SoIS. We emphasize that SoIS is oriented towards inter-organizational business processes, which intensily impacts on the notations to support their specification and in their engineering itself. As such, SoIS has the potential to trigger the emergence of and also establish inter-organizational business process (and associated interoperation) among the constituents to achieve the SoIS goals. \m{In this scenario, we introduced the concept of Process-of Business Processes \cite{Cagnin2021a}, which brings a new understanding of the business processes associated with SoIS and necessarily required new means to manage such complex, dynamic processes \cite{Cagnin2021c}.}

%\subsection*{Q4: Why is SoIS necessary as a particular type of system? \\ Q5: What are the engineering needs to be raised by SoIS as a particular subtype of SoS?}

\subsection{\textbf{\m{Q4: Why is SoIS necessary as a particular type of system? and Q5: What are the engineering needs to be raised by SoIS as a particular subtype of SoS?}}}

In short, SoIS is needed as a particular 
%class 
\m{type} of SoS to motivate advances that SoS notations and tools have not coped with. We discuss it more in-depth in Section~\ref{s:gaps-challenges}, where we discuss gaps and research opportunities.

\section{Discussion}
\label{s:discussion}
After answering the aforementioned questions, further issues could still arise. We use some motivating scenarios for discussing them.

Some facts motivate the study of SoIS as a particular  
%class 
\m{type} of systems. Maier \cite{maier_architecting_1998} argues in his seminal paper that ``what justifies the creation of a new class of systems is the insufficiency of the existing techniques to deal with the particularities of those emerging systems''. %Operational and managerial independence of systems were a novelty that justified the creation of the new term at that time (System-of-Systems).
For SoIS, we claim that the presence of IS is the particularity that makes the state-of-the-art techniques for SoS not being sufficient, in particular\m{, concerning the specific means to design, implement, and evolve them, including the tigh relationship between technical and business levels.} 
%the specification notations. %imposes the establishment of business processes that can not be precisely documented using the state of the art; as a result, the SoIS engineering itself could be any type of software-intensive system able to establish interoperability links has the potential to be part of a SoS or form a new subtype of SoS. Hence, we could observe the emergence of SoS majorly formed by Artificial Intelligence (AI) software systems (Intelligent SoS/SoIS), Robotic SoS, Blockchain SoS, and other types; (iii) we are already aware that some of the state-of-the-art modeling notations and tools such as mKAOS \cite{silva2015}, UML, SySML and BPMN  do not support a precise specification of flexible and inter-organization business processes. This justifies the investigation of this class of systems in the forthcoming years;

\begin{figure*}[!ht]
\centering
\includegraphics[width=5.0in]{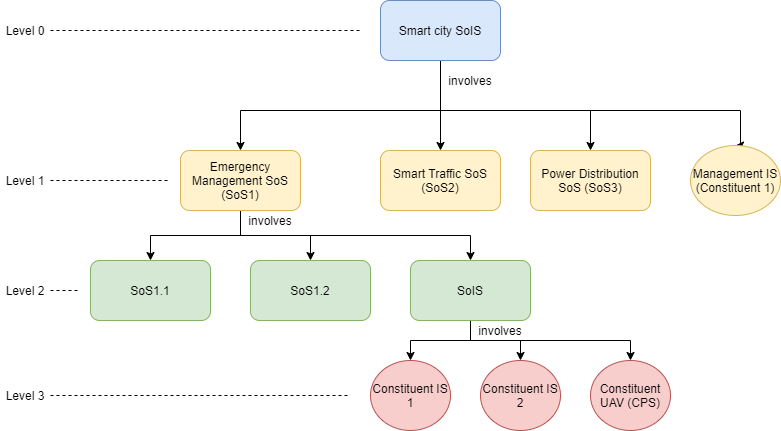}
% suggested caption for Figure 2: Ahmad 
\caption{Smart city SoIS hierarchy: SoS, SoIS with IS and Non IS/CPS constituents.}
\label{fig_2}
\end{figure*}

From a hierarchical point of view, a smart city SoIS can be composed of several Emergency Management SoS (SoS1), Smart Building SoS (SoS2), a Smart Traffic SoS, and Power Distribution SoS, besides a different management IS (Constituent IS 1). 
%However, an 
\m{An} Emergency Management SoS (SoS1) can also be itself composed of a SoIS as a constituent. As illustrated in Fig. \ref{fig_2}, an SoS can be composed of other inner SoS and SoIS. We argue that an SoS is only considered a SoIS when it directly contains one or more IS in its 
%constituent set, 
\m{set of constituents,} i.e., in its immediate hierarchy level. From Fig. \ref{fig_2}, we can say that SoS1 is not a SoIS, although one of its constituents is itself a SoIS. The rationale for this decision is that an SoS should contain one or more IS 
%in its constituents set , 
to be considered a SoIS. If the IS is not in that hierarchical level, then the SoS should not be treated as a SoIS since it will not demand SoIS engineering needs and it will not involve a business process to interoperate the constituents, which is the colorary of characteristic 2 (i.e., goals expressed as a business process, which is not mandatory, but important). %since information information in that type of SoS is often aggregated and analyzed in an upper instance. The constituent SoIS (Level 2) can only deliver information that can flow over the SoS1 architecture, not necessarily using a business process. Moreover, even if this happens,

%FIGURE 2

\section{Research Gaps and Challenges}
\label{s:gaps-challenges}
SoIS engineering is still a maturing discipline. As such, 
%some 
gaps and research challenges 
%still 
remain 
%and we discuss them in the following.
\m{as discussed below.}

%Some studies 
\m{Studies} in the literature %\cite{Teixeira2019,gracianoNeto2018SBSI,Fernandes2019,Neves2020}
revealed that \m{\cite{Teixeira2019,gracianoNeto2018SBSI,Fernandes2019,Neves2020}}: (i) at least three different definitions for SoIS were available in the literature, thus being not consensual yet; %(ii) the research on SoIS was majorly conducted over the years under two perspectives: an architecting perspective (leaded by our group) and an IS perspective, where SoIS is a parallel between SoS and IS and a SoIS is seen as a IS in which components are themselves systems \cite{saleh2015information}, (iii) SoIS have been studied and applied in 12 different applications domains, ranging from education to government, health and smart cities, 
(ii) there was a lack of a common sense on a set of characteristics that determine what is a SoIS and what is not; and (iii) the main problems introduced by SoIS and not solved by SoS until now are majorly on modeling and, by extension, its engineering.

\textbf{SoIS specification.} Modelling is an inherent part of Systems Engineering. A system model should provide a precise (even though abstract) representation of its counterpart in reality. mKAOS \cite{silva2015}, the state-of-the-art notation for modeling SoS missions/goals, does not support the notion of sequence %between 
\m{among} activities, which is essential to model SoIS goals. For instance, it is important to represent a constituent held by a %company 
\m{organization/entity} solving a portion of a global mission and forwarding data or the ``control flow'' for another following constituent to go ahead until concluding the SoIS global goal. Even the languages that could support such representation\m{,} such as UML\footnote{https://www.omg.org/spec/UML}, SySML\footnote{https://www.omg.org/spec/SysML}\m{,} and BPMN\footnote{https://www.omg.org/spec/BPMN/2.0/}\m{,} do not support a precise specification of SoS/SoIS dynamic architectures and the multiple systems and organizations\m{/entities} involved (they were designed to represent single systems and single enterprises). We need modeling languages that match SoIS requirements so that we can have supporting tools and processes that work well for precisely capturing SoIS architectures. BPMN seems to be the best candidate for modeling at least the ``process view'' 
%of such an architecture 
\cite{gracianoNeto2018SBSI}\m{\cite{Cagnin2021c}}. However, it still needs to evolve towards coping with those requirements (dynamic architecture and native support for multiple organizations\m{/entities}).%and systems since BPMN currently supports the representation of processes for a single organization and a static process. %We claim that the business processes arising in a SoIS are potentially flexible (they can change over time), inter-organizational (involve several companies) and should be managed and executed (using executable languages). Then, there are multiple requirements that should be covered so that a language could precisely support capturing SoIS missions and architecture. After such language emerge, SoIS engineering could really succeed  

\textbf{How embracing are SoIS compared to SoS?} To formalize part of the proposed concepts, let us consider that each SoS proposed, planned, specified or engineered so far are represented in a mathematical set ($S_{SoS}$). A question to raise is: how large is the SoIS subset ($S_{SoIS}$) compared to SoS?, i.e., how many existent/proposed SoS have some IS involved among their constituents? Firstly, we state that $S_{SoS} \supset S_{SoIS}$. We already know that some SoS are not SoIS, such as Flood Monitoring SoS, CPSoS in general, and a Smart Building \cite{GracianoNetoiSyS}, i.e., $\exists s \in  S_{SoS} \mid s \notin S_{SoIS}$. One can argue that if the cardinality of SoIS is close to SoS, then most of the SoS are SoIS and perhaps the existence of SoIS as a particular 
%class 
\m{type} of system could be questioned. Further research should be conducted to answer this question.

%\textbf{Absence of notations and tools to deal with SoIS:} Modelling is an inherent part of systems engineering. A system model should provide a precise (although abstracted) representation of its counterpart in reality. However, SoS (and SoIS) inherently have dynamic architectures. None of the existing modeling notations support the specification of constituents joining and leaving the SoS at runtime (even this event occurs not so frequently) yet. A study has revealed that the most suitable language to be evolved to cope with SoIS requirements is BPMN due to its acceptance in the industry, widespread dissemination, and native support for business processes representation \cite{gracianoNeto2018SBSI}. Hence, notations should be evolved/proposed, and tools should also be created to support better SoIS engineering, including executable ones \cite{Lebtag2020}. 
%\textbf{SoIS missions specification repositories:} Over the years studying SoIS, we have observed that authors have not made available the SoS/SoIS documentation in papers and other scientific repositories. Hence, we do not have the raw material to assess SoIS goals specification empirically. We claim the need for a SoIS documentation repository. Henceforth, authors can make available their SoIS architectural documentation, particularly the missions/goals specification, which are essential to investigate the suitability of available notations for SoIS specification and propose evolutions to cope with SoIS needs.
% Suggested: Ahmad 

\textbf{Modeling and simulation for SoIS.} 
Although there has been extensive work on SoS software architecture modeling, verification, and simulation over the years \cite{Cadavid2020}, the same has not happened to SoIS. SoIS needs specific formal modeling and verification techniques to deal with its underlying complexity\m{/dynamism, mainly that associated with the business process level} \cite{guessi2019architectural}\m{\cite{Cagnin2021a}}. At an abstract level, SoIS architectural models should have the capability to describe structures and behaviors with suitable notations for achieving goals with the collaborative IS constituents and dealing with organizational and business processes needs. As a SoIS has a heterogeneous set of consitutent systems, formal models with well-defined architectural styles/patterns can address interoperability for better exchange of information among diverse constituent systems. The resulting architectural models should be executable to validate emergent behaviors and conformance of core quality attributes of SoIS with formal verification and simulation tools coupled with model-driven engineering. 

\textbf{Implementation of real-world SoIS.} We still observe the absence of more complete implementations of real-world SoIS. Saleh 
%and colleagues~
\m{et al.} \cite{saleh2015information} 
%have 
described the implementation of a SoIS in the educational domain\m{, while}
%. 
Neves et al. \cite{Neves2020} %have also implemented 
\m{presented} a prototype of a controller for a directed SoIS\m{, but}
%. However, 
SoIS in other \m{diverse} domains should also be implemented\m{.}
%and the domains should be diversified. 
Overcoming this challenge is inherently related to overcome other challenges, such as\m{:} (i) the full interoperability for constituents \m{\cite{Cagnin2021a}}\cite{fullinteroperability}, (ii) the SoIS dynamic architectures \cite{Manzano2020}, and (iii) challenges related to operational and managerial independence \cite{teixeira_constituent_2020}. For achieving it, we could use an arrangement of IS from organizations\m{/entities} with the lenses of SoS to discover and balance business processes (SoIS and constituent IS) that could be established among them, besides analyzing and involving the corresponding IT architecture for means of interoperation\m{~\cite{Cagnin2021a}}. %There are still important advances until reaching a plain real SoIS and the scientific community can invest effort on it in the forthcoming years.

\section{Final Remarks}
\label{s:finalRemarks}
The main contribution of this paper was to compile, summarize, and establish an understanding 
%on 
of Systems-of-Information Systems (SoIS). The content of this paper is the product of a five-year research project on SoIS developed in cooperation between multiple institutions in Brazil with international cooperations in Australia, \m{the} USA, and France. We remark that  SoIS has been recognized as a promising and prominent research area and listed as one of the Grand Challenges in Information Systems for the Decade 2016-2026 in Brazil \cite{neto2017smart}. As CPSoS has evolved over the past years to become an important area of interest for the SoSE community, we also expect that SoIS can also reach such status in the forthcoming years, attracting researchers and contributors. We expect this paper 
%to work 
\m{works} as a theoretical milestone to 
%support 
\m{motivate} newcomers and experienced researchers %in that direction.
\m{to futher consolidate a SoIS international community.}

\bibliographystyle{IEEEtran}
\bibliography{refs}
\end{document}